\documentclass[sigconf]{acmart}

\usepackage{booktabs} % For professional-looking rules (\toprule, \midrule, \bottomrule)
\usepackage{float} 
\usepackage{array} 
\usepackage{multirow}
\usepackage{tabularx} % For tables that automatically fit a specific width
\usepackage{color}
\usepackage{subcaption}
\usepackage{xspace}
\usepackage{enumitem}
\usepackage{soul}
\usepackage{listings}
\definecolor{burntorange}{rgb}{0.8, 0.33, 0.0}
\definecolor{byzantium}{rgb}{0.44, 0.16, 0.39}
\definecolor{byzantine}{rgb}{0.74, 0.2, 0.64}
\definecolor{lightlightgray}{rgb}{0.94, 0.94, 0.94}
\definecolor{lightblue}{rgb}{0.91, 0.95, 0.99}

\AtBeginDocument{%
  }

\copyrightyear{2026}
\acmYear{2026}
\setcopyright{cc}
\setcctype{by}
\acmConference[CHI '26]{Proceedings of the 2026 CHI Conference on Human Factors in Computing Systems}{April 13--17, 2026}{Barcelona, Spain}
\acmBooktitle{Proceedings of the 2026 CHI Conference on Human Factors in Computing Systems (CHI '26), April 13--17, 2026, Barcelona, Spain}
\acmPrice{}
\acmDOI{10.1145/3772318.3791158}
\acmISBN{979-8-4007-2278-3/2026/04}

\newcommand{\hlc}[2][yellow]{{%
    \colorlet{foo}{#1}%
    \sethlcolor{foo}\hl{#2}}%
}

\newcommand{\hotkeyName}[1]{\hlc[lightlightgray]{\normalsize\texttt{#1}}}

\begin{document}
\renewcommand{\arraystretch}{1.3}
\title{ADCanvas: Accessible and Conversational Audio Description Authoring for Blind and Low Vision Creators}


\author{Franklin Mingzhe Li}
\affiliation{%
  \institution{Carnegie Mellon University}
  \city{Pittsburgh}
  \state{PA}
  \country{United States}
}
\affiliation{%
  \institution{Google Research}
  \city{New York}
  \state{NY}
  \country{United States}
}
\email{mingzhe2@cs.cmu.edu}

\author{Michael Xieyang Liu}
\affiliation{%
  \institution{Google DeepMind}
  \city{Pittsburgh}
  \state{PA}
  \country{United States}
}
\email{lxieyang@google.com}

\author{Cynthia L. Bennett}
\authornote{These authors jointly supervised this work.}
\affiliation{%
  \institution{Google Research}
  \city{New York}
  \state{NY}
  \country{United States}
}
\email{clbennett@google.com}

\author{Shaun K. Kane}
\authornotemark[1]
\affiliation{%
  \institution{Google Research}
  \city{Boulder}
  \state{Colorado}
  \country{United States}
}
\email{shaunkane@google.com}

\renewcommand{\shortauthors}{Li et al.}

\begin{abstract}
Audio Description (AD) provides essential access to visual media for blind and low vision (BLV) audiences. Yet current AD production tools remain largely inaccessible to BLV video creators, who possess valuable expertise but face barriers due to visually-driven interfaces. We present ADCanvas, a multimodal authoring system that supports non-visual control over audio description (AD) creation. ADCanvas combines conversational interaction with keyboard-based playback control and a plain-text, screen reader–accessible editor to support end-to-end AD authoring and visual question answering (VQA). Combining screen-reader-friendly controls with a multimodal LLM agent, ADCanvas supports live VQA, script generation, and AD modification. Through a user study with 12 BLV video creators, we find that users adopt the conversational agent as an informational aide and drafting assistant, while maintaining agency through verification and editing. For example, participants saw themselves as curators who received information from the model and filtered it down for their audience. Our findings offer design implications for accessible media tools, including precise editing controls, accessibility support for creative ideation, and configurable rules for human-AI collaboration.
\end{abstract}

\begin{CCSXML}
<ccs2012>
   <concept>
       <concept_id>10003120.10011738.10011776</concept_id>
       <concept_desc>Human-centered computing~Accessibility systems and tools</concept_desc>
       <concept_significance>500</concept_significance>
       </concept>
 </ccs2012>
\end{CCSXML}

\ccsdesc[500]{Human-centered computing~Accessibility systems and tools}

\keywords{Accessibility, Audio Description, Conversational Agent, MLLM, Blind, Low Vision, AD Creator}
\begin{teaserfigure}
  \centering
  \includegraphics[width=0.85\textwidth,
  alt={Screenshot of the ADCanvas interface with five labeled sections. On the left, section (a) “Video” shows a list of video titles and a preview window paused at 0:35 with two eggs frying in a pan. Below it, section (b) “Hotkeys” lists keyboard shortcuts such as “Ctrl+/ Describe a high-level summary of the video” and “Ctrl+1 Play/Pause Video.” Beneath that, section (c) “Conversation History” displays a chat between the agent and user: the agent describes a scene of a man smiling at fried eggs forming a face in the pan, and confirms it was inserted into the script; a text input box (C1) is available for commands. On the right, section (d) “Audio Description” shows a script with timestamped narration, such as “0 min 2 sec to 0 min 7 sec The alarm clock rings loudly. A man wakes up in bed, looking distressed.” One portion (d2) is enlarged for editing, with text: “0 min 33 sec to 0 min 38 sec. Two fried eggs in a pan form a smiling face. The man smiles.” Editing controls (d3) allow applying changes. At the bottom, overlay (d1/e) displays the same segment highlighted, labeled “0 min 33 sec to 0 min 38 sec.” Labels a–e mark each feature area.}]{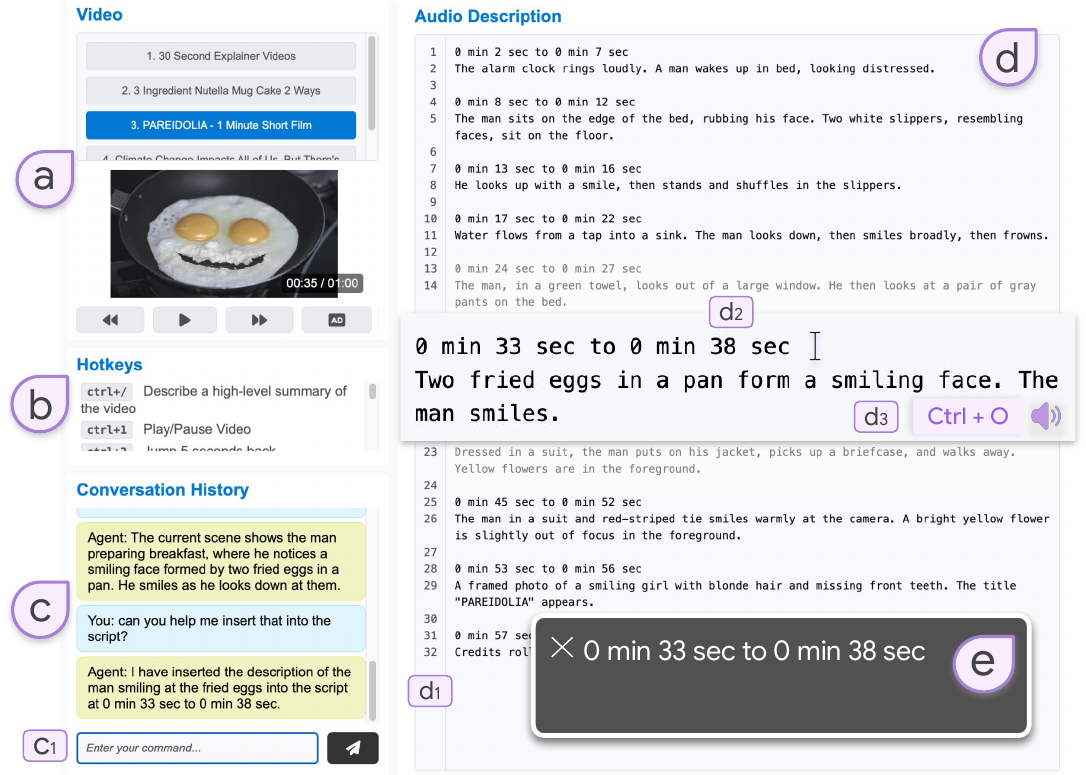}
  \vspace{-2mm}
  \caption{ADCanvas allows the user to control the video (a) through both screen reader compatible buttons (a) and hotkeys (b). The user can communicate with the conversational agent through the textbox (c1) to ask questions or manipulate the AD script (c, d). The user can also edit the AD script through the WebVTT-based AD editor (d, d2) line by line (d1) with a screen reader (e). Upon completing the AD scripts, the user can play the AD narration with the original soundtrack by using hotkey (d3).}
  \Description{Screenshot of the ADCanvas interface with five labeled sections. On the left, section (a) “Video” shows a list of video titles and a preview window paused at 0:35 with two eggs frying in a pan. Below it, section (b) “Hotkeys” lists keyboard shortcuts such as “Ctrl+/ Describe a high-level summary of the video” and “Ctrl+1 Play/Pause Video.” Beneath that, section (c) “Conversation History” displays a chat between the agent and user: the agent describes a scene of a man smiling at fried eggs forming a face in the pan, and confirms it was inserted into the script; a text input box (C1) is available for commands. On the right, section (d) “Audio Description” shows a script with timestamped narration, such as “0 min 2 sec to 0 min 7 sec The alarm clock rings loudly. A man wakes up in bed, looking distressed.” One portion (d2) is enlarged for editing, with text: “0 min 33 sec to 0 min 38 sec. Two fried eggs in a pan form a smiling face. The man smiles.” Editing controls (d3) allow applying changes. At the bottom, overlay (d1/e) displays the same segment highlighted, labeled “0 min 33 sec to 0 min 38 sec.” Labels a–e mark each feature area.}
  \label{fig:teaser}
\end{teaserfigure}

\maketitle

\section{Introduction}
Audio Description (AD) is a critical access modality that augments visual media content with spoken narration for people who are blind or have low vision (BLV)~\cite{snyder2008audio, jiang2022co, cheema2025describe, natalie2024audio}. Far beyond an auxiliary feature, AD is a primary means through which BLV audiences access film, television, and user-generated videos posted on social media~\cite{jiang2024s, hattich2020hear}. Narration is inserted into gaps with no dialogue or important sounds, and might describe a character's critical facial expression, the intricate details of a fantasy setting, or the key actions in a fast-paced sequence that dialogue alone cannot capture, making content comprehensible and engaging.

A paradox about AD is that some of its most skilled practitioners—scriptwriters, narrators, and quality reviewers—are themselves BLV~\cite{ReidMyMi63:online}, yet they routinely face barriers when working with AD creation tools that are fundamentally designed for sighted users \cite{Describi84:online}. Mainstream AD authoring tools, such as Digital Audio Workstations (DAWs) and non-linear video editors, depend on highly visual metaphors like timelines, waveform editors, and drag-and-drop interfaces~\cite{braun2007audio}. This design hegemony renders the tools difficult or impossible to use independently with screen readers, often forcing BLV creators into inefficient workarounds or dependence on sighted collaborators~\cite{jiang2022co, braun2007audio}. Moreover, BLV creators lack robust visual question answering (VQA) systems that are integrated into AD authoring workflows \cite{jiang2022co}, often forcing them to rely on external tools or sighted collaborators for visual information. This fragmented workflow reduces BLV creators’ direct control over what visual details are considered during authoring. As a result, sighted collaborators may not prioritize the same visual information as a BLV creator, leading to AD that omits subtle non-verbal cues or reflects different scene interpretations, which can fundamentally alter the quality and authenticity of the final AD.\looseness=-1

While prior HCI and accessibility research has addressed certain AD development pain points (e.g., fitting lines of AD between dialogue), these efforts often target sighted users~\cite{pavel2020rescribe, YouDescr90:online, cheema2025describepro} and may exclude BLV creators. This accessibility gap raises the question: \textit{how might AD authoring tools be reimagined to support creative workflows for BLV AD creators?} An answer to this question may lie in the emergence of multimodal Large Language Models (MLLMs), which offer an opportunity to rethink accessible media creation~\cite{alayrac2022flamingo, wang2024lave, li2022feels, cheng2025text, li2022exploration, xiao2025understanding,li2025oscar,li2021non,li2024recipe}. Given the ability of MLLMs to interpret and translate visual inputs \cite{team2023gemini,liu2025gensors,liu2025meta,olwal2025semantic,zhang2025vizxpress,ning2025aroma,li2025exploring}, they may be useful in supporting AD creation. Furthermore, these systems can combine natural language input and context awareness to support users in performing tasks through conversation with an agent~\cite{wang2020human, wermelinger2023using,liu2023wants,petridis2026compass,qian2025llm,li2023understanding}. Such systems may enable \textit{co-creative workflows} \cite{petridis2024situ} in which BLV users drive the authoring process with an AI as a supporting partner.

To explore this new paradigm of conversational co-creating AD, we introduce ADCanvas\footnote[1]{ADCanvas stands for ``Audio Description Collaborative Authoring for Narrative Video Accessibility Synthesis''} (Figure \ref{fig:teaser}), a novel multimodal AD authoring tool for BLV AD Creators. ADCanvas explores how the AD creation process can be reconceptualized to include BLV creators. ADCanvas enables users to generate and revise AD scripts through a combination of a conversational AI agent and accessible keyboard navigation. It combines VQA, script generation and refinement, hotkey-based video control, and a WebVTT-based script editor. Using ADCanvas as an exploratory probe with BLV AD creators, we explored the following research questions:

\begin{itemize}
    \item RQ1: How does collaboration with an embedded multimodal agent shape the practices and creative workflows of BLV AD creators?
    \item RQ2: What are the key interaction design challenges and opportunities in creating non-visual and conversational workflows for complex creative tasks like AD authoring?
    \item RQ3: What are BLV creators' perceptions of human-AI collaborative workflows for AD authoring?
\end{itemize}

We conducted a qualitative user study with 12 BLV participants, including both professional AD practitioners and video creators, who used ADCanvas to author AD scripts for short videos. We found that: (1) the conversational paradigm supported an accessible AD-creation process; (2) our participants guided the AI through creative interaction patterns (e.g., using AI to create first drafts or drafting text manually with VQA support from AI); (3) participants maintained a supervisory stance with the agent that balanced trust with verification; and (4) participants’ experiences revealed usability breakdowns and opportunities related to agency, precision, and interaction fluidity. Participants enjoyed collaborating with ADCanvas’s AI agent, expressing a strong desire to continue using it after the study. In discussing their use of ADCanvas, participants considered the system a responsive co-author that enabled them to exercise independent, high-quality creative judgment while receiving appropriate levels of accessibility support.

In summary, our contributions are:

\begin{itemize}[leftmargin=0.2in]
    \item We introduce ADCanvas, a novel, screen reader-accessible multimodal AD authoring tool which enables BLV creators to generate and revise AD scripts through contextual and conversational interaction, VQA, keyboard navigation, and real-time in-line narration.
    \item We provide empirical findings from a study with 12 BLV participants, highlighting the practices of human-AI co-creation, the negotiation of trust and verification, and the breakdowns in current conversational workflows.
    \item We derive design implications for future accessible creative tools, emphasizing agent configurability and fine-grained authoring control for BLV AD creators.
\end{itemize}

\section{Related Work} 
This section first deconstructs the sociotechnical barriers in conventional AD tools that hinder BLV creators. We then examine how semi-automated and community-driven platforms address aspects of AD production but leave core interactional barriers unresolved. Finally, we review prior work on instruction-based LLM agents for AD creation.

\subsection{Professional Media Creation Tools and BLV Creators}
The ``gold standard'' for professional AD creation is a manual process that demands both artistic interpretation and technical precision~\cite{snyder2008audio, braun2007audio}. To support this, practitioners rely on specialized software, such as Ooona or Subtitle Edit Pro~\cite{Subtitle50:online, AllMedia24:online,ncam_cadet}. The design philosophy of these tools, however, is deeply rooted in a visual-centric paradigm. For example, when editing a line of AD, the user interacts with a graphical timeline where video and audio are shown as horizontal bars~\cite{campos2020cinead}. To insert a description like ``a woman in a red coat glances at her watch,'' the user finds a pause in the dialogue and drags on the timeline to set the in- and out-points~\cite{campos2020cinead}. To verify sync with on-screen action, they “scrub” a playhead for frame-by-frame adjustment. Audio mixing, crucial for balancing AD with original sound, is done by manipulating faders, panning knobs, and drawing volume automation curves on the audio tracks. This visual-spatial paradigm is fundamentally incompatible with the linear, serialized output of screen readers, making it hard to build a mental model of concurrent events and timings and rendering inaccessible many cues and operations that define professional AD editing, including waveform contours, timing gaps, layered audio tracks, automation envelopes, and frame-accurate scrubbing, none of which translate into a linear auditory stream \cite{villena2014web}. Such visual dependency renders core authoring tasks inaccessible to BLV creators, who cannot independently navigate timelines, identify gaps, or adjust timing~\cite{villena2014web}.

In response to this technological gap, BLV AD creators collaboratively work with a sighted partner who answers visual questions when creating AD \cite{bennett2025made, jiang2023beyond,jiang2022co,hirvonen2023guided}. This human partnership can yield high-quality results, synergistically combining different skill sets to navigate the limitations of the software. However, this collaborative model can introduce logistical overhead, require synchronous work, and tether the creative flow of the BLV creator to the availability of their sighted partner~\cite{hirvonen2023guided}. To this end, we explore design spaces of systems that would increase the creative autonomy of BLV creators. ADCanvas enables tasks that existing visual tools cannot currently support non-visually, including independent visual understanding through conversational VQA, accessible script drafting, and non-visual navigation of timestamped AD content.

\subsection{Platforms for Collaborative and Community-Based AD}
\subsubsection{AI Assistance for Content Generation and Spotting}
A major focus of HCI and AI research has been on leveraging artificial intelligence to alleviate the cognitive and manual burden of writing AD scripts. This has led to a class of semi-automated, Human-in-the-Loop (HITL) systems where AI performs initial analysis and a human provides refinement and oversight~\cite{yuksel2020human, zhang2023imageally,do2025youdescribe}. For example, systems can use automatic silence detection to suggest potential ``spots'' for AD insertion~\cite{lakritz2006semi} or, more recently, use powerful Vision-Language Models to generate draft descriptions for entire scenes~\cite{alayrac2022flamingo, krishna2017dense}.\looseness=-1

These AI-generated suggestions are then presented to a human for review and editing. However, these systems primarily serve sighted people as creators and BLV people as consumers. Systems like \textit{Rescribe} and \textit{DescribePro} were designed to make the professional workflow more efficient for a sighted author, who could visually review AI suggestions and then use a standard timeline interface to make final adjustments~\cite{pavel2020rescribe,cheema2025describepro}. Fully automated systems are primarily a mechanism for providing consumers with rudimentary, on-the-fly descriptions where no professional AD exists, with quality and narrative coherence being secondary concerns~\cite{natalie2023supporting, ning2024spica, han2023autoad, wang2021toward,stangl2023potential}. The fundamental interaction model for how and when to precisely place, edit, and mix the description remains inaccessible to BLV AD creators. Our work aims to address this gap by providing non-visual control over the entire scripting process.\looseness=-1

\subsubsection{Crowdsourced and Volunteer-Based AD Platforms}
A successful strategy for increasing AD’s prevalence has been harnessing community collaboration. Killough and Pavel explored community AD for livestreams \cite{killough2023exploring}, And YouDescribe, a landmark project, provides a web-based platform where volunteers crowdsource AD for YouTube videos~\cite{YouDescr90:online}. Recently, YouDescribe was augmented with AI-generated descriptions, making the HITL process more inclusive by allowing BLV viewers to rank AD \cite{do2025youdescribe}. By distributing the labor of AD creation, such platforms have made significant progress in addressing the AD gap for popular content. They serve as key infrastructures for community organization and content sharing. However, demand still far exceeds what volunteers can produce, partly due to the time, training, and confidence required~\cite{stangl2023potential}. Moreover, these platforms often replicate the standard visual timeline metaphor, limiting accessibility for BLV individuals who wish to contribute. Our work on ADCanvas is complementary, focusing on a core authoring engine that could be integrated into such platforms to support both sighted and BLV creators.\looseness=-1

\subsection{Instruction-based LLM Agents for AD Creation}
Prior research leveraged the ability of Large Language Models (LLMs) to be steered through direct, natural language instructions \cite{wei2021finetuned,ouyang2022training}. Recent HCI work has demonstrated that LLM-based systems can be guided by users' instructional inputs~\cite{liu2023wants,petridis2024constitutionmaker,liu2024we}. Studies have shown that instruction-following LLMs can incorporate iterative feedback to refine their outputs, such as conversational code generation~\cite{liu2023wants,petridis2024constitutionmaker}. Other systems explore how high-level natural language commands can be translated into specific system actions, allowing for more intuitive control over complex processes~\cite{petridis2024situ}. This body of work shows a trend toward enabling users to shape and direct AI behavior through conversational interaction, rather than relying on graphical interfaces or rigid programming. 

This conversational, instruction-based paradigm is well-suited to AD scripting. The practice of AD is already governed by a set of established professional guidelines and best practices (e.g., ``use present tense,'' ``be objective, do not interpret,'' ``prioritize describing actions over settings'')~\cite{snyder2008audio}. A conversational agent can be explicitly instructed to adhere to these foundational rules. More importantly, it can allow a BLV creator to instill their own expert knowledge and stylistic preferences by providing iterative feedback and direct instructions, such as ``be more concise in the next description'' or ``focus on the character's facial expression.'' This approach offers a mechanism for fine-grained, non-visual control that is missing from current tools. While some systems are emerging with features like VQA, they often lack a structured method for users to instill and iteratively refine the agent's behavior through conversation~\cite{stangl2023potential}.

ADCanvas is designed as a technology probe to investigate this conversational paradigm for accessible media creation. It explores how a conversational agent, in tandem with hotkeys, an accessible editor, and a UI optimized for screen reader navigation, can empower BLV users to not only generate AD content but also to progressively shape the agent's behavior to match their expert knowledge and creative intent, effectively building a personalized AD creation assistant.

\section{The ADCanvas System}
\label{sec:system}

ADCanvas is an accessible AD editor that empowers BLV creators to independently author AD for videos, by integrating state-of-the-art MLLMs into a novel, context-aware workflow. ADCanvas exposes two linked but separable surfaces, a plain-text WebVTT editor and a set of keyboard-based media controls. The conversational panel allows users to ask visual questions and to generate script drafts. ADCanvas allows creators to create and edit AD scripts using a screen reader (e.g., JAWS~\cite{freedomscientific_jaws}, NVDA~\cite{NVAccess81:online}, VoiceOver~\cite{Accessib64:online}), controlling video playback, asking visual questions, editing the AD script, and controlling the system via natural language (Figure \ref{fig:teaser}). The core contribution of ADCanvas is removing visual dependencies of AD creation to support creative synthesis and narrative construction. At the same time, ADCanvas is not intended to replace professional DAWs; some advanced operations supported by commercial tools, such as fine-grained timing alignment or waveform-based gap detection, remain outside its current scope. Accordingly, we position ADCanvas as a technology probe \cite{hutchinson2003technology} that explores how multimodal agents can expand BLV creators’ access to AD authoring workflows, rather than as a full substitute for industry-standard production environments.

\subsection{Example Usage Scenario}
We illustrate the experience of using ADCanvas by describing an example usage scenario that embodies many of the use cases drawn from our later user study. While this is one way to use the system, there is flexibility in how it is used. 

Clara, a blind AD creator and screen reader user, is tasked by a client with creating an AD track for a short film, ``PAREIDOLIA\footnote[1]{https://www.youtube.com/watch?v=JLmOkEEC9SQ}.’’ She opens the video file in ADCanvas.

To begin, Clara wishes to get an overview of the film. She uses the playback hotkeys (\hotkeyName{Ctrl + 1} to play/pause, \hotkeyName{Ctrl + 2} to rewind) to perform a full pass-through, listening to the music to understand the film's pacing and emotional tone. To understand the overall video summary, she navigates to the conversational agent (Figure \ref{fig:teaser}-c1) and then: \\

\begin{figure}[t]
    \centering
    \includegraphics[width=0.8\columnwidth,alt={Screenshot of a conversation interface between a user and an AI agent. The dialogue is displayed in colored boxes, with user messages in blue and agent responses in yellow. In the first exchange (a1), the user asks: “Provide a brief summary of the entire video, focusing on the main character and any recurring visual themes.” The agent replies (a2): “The video’s main character is a man who initially appears distressed upon waking but finds moments of subtle joy throughout his morning routine. A recurring visual theme is …” In the second exchange (b1), the user asks: “What visuals show the man is looking distressed?” The agent responds (b2): “The man appears distressed in several instances. Initially, upon waking around 0 minutes 2 seconds, his eyes are closed tightly, and when he opens them, his expression is weary and sad. At 0 minutes 7 seconds, he sits up and covers his face with his hands …” Labels a1, a2, b1, and b2 mark each user or agent message.}]{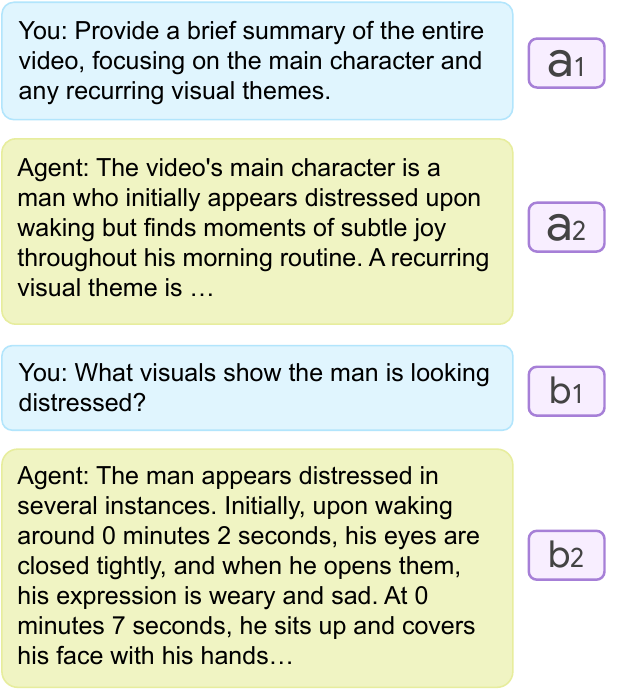}
    \vspace{-2mm}
    \caption{Conversations between Clara and the agent when asking details about the “PAREIDOLIA” video. a1) Clara asked the agent to provide a brief summary. a2) The agent responded with text showing the man appears distressed. b1) Clara continued to ask what visuals show the man is looking distressed. b2) The agent tells details regarding where and why the man looks distressed.}
    \label{fig:conversation}
    \Description{Screenshot of a conversation interface between a user and an AI agent. The dialogue is displayed in colored boxes, with user messages in blue and agent responses in yellow. In the first exchange (a1), the user asks: “Provide a brief summary of the entire video, focusing on the main character and any recurring visual themes.” The agent replies (a2): “The video’s main character is a man who initially appears distressed upon waking but finds moments of subtle joy throughout his morning routine. A recurring visual theme is …” In the second exchange (b1), the user asks: “What visuals show the man is looking distressed?” The agent responds (b2): “The man appears distressed in several instances. Initially, upon waking around 0 minutes 2 seconds, his eyes are closed tightly, and when he opens them, his expression is weary and sad. At 0 minutes 7 seconds, he sits up and covers his face with his hands …” Labels a1, a2, b1, and b2 mark each user or agent message.}
    \vspace{-3mm}
\end{figure}

Clara asks: \texttt{``Provide a brief summary of the entire video, focusing on the main character and any recurring visual themes''} (Figure \ref{fig:conversation}-a1). 

The agent responds in a few seconds: \texttt{``The video's main character is a man who initially appears distressed upon waking but finds moments of subtle joy throughout his morning routine. A recurring visual theme is…''}  (Figure \ref{fig:conversation}-a2). \\

This confirms her initial impression from the music and gives her a narrative foundation for the AD script.\\

Clara asks for more detail: \texttt{``What visuals show the man is looking distressed, are there any actions that the man made?''} (Figure \ref{fig:conversation}-b1) 

The agent provides a factual description: \texttt{``...at 0 minutes 8 seconds, he sits up and covers his face with his hands''} (Figure \ref{fig:conversation}-b2).\\

With this foundational understanding, Clara is ready to structure the script. To start, she issues a direct command to the agent: \texttt{``Identify all the silent gaps available for audio descri- ption and generate the timestamps and audio descriptions.''} In about ten seconds, ADCanvas processes the video's audio track and populates the script editor with AD script: a list of timestamped entries, each followed by a suggested line of AD (Figure \ref{fig:teaser}-d). 

\begin{figure}[t]
    \centering
    \includegraphics[width=0.9\columnwidth,alt={Screenshot of the ADCanvas interface showing a video frame, transcript, and conversation. At timestamp 0:08, the video (label 2) shows a man sitting on the edge of his bed, rubbing his face, with two white slippers on the floor. The transcript panel (label 1) reads: “0 min 8 sec to 0 min 12 sec. The man sits on the edge of the bed, rubbing his face. Two white slippers sit on the floor.” In the conversation panel, the user asks (label 3), “Anything special about the slippers?” and the agent replies in a yellow box, “Yes, it appears that the white slippers resemble smiling faces at 0:10.” A small thumbnail (label 3a) below shows a close-up of the slippers, which resemble smiling faces. Labels 1, 2, 3, and 3a identify the transcript, timestamped video, user–agent conversation, and supporting thumbnail.}]{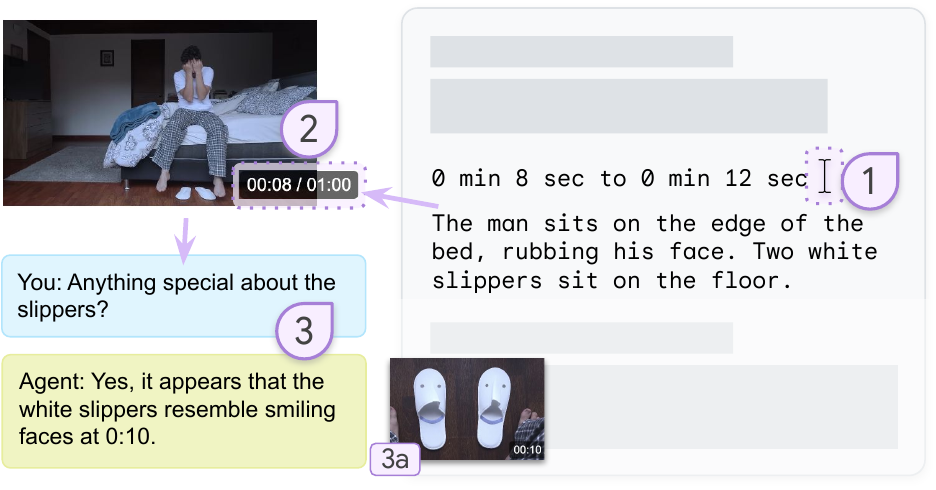}
    \caption{Context-aware interaction between the AD editor, video trigger, and the conversational agent. 1) The user selects the line starting from 0 minute 8 seconds. 2) The video cursor directly jumps to the start time of this scene. 3) The user asks the agent anything special about the slippers contextually. 3a) Already made aware of the timestamp that the user is focused on, the agent responded with the corresponding details of the slippers.}
    \label{fig:contextaware}
    \Description{Screenshot of the ADCanvas interface showing a video frame, transcript, and conversation. At timestamp 0:08, the video (label 2) shows a man sitting on the edge of his bed, rubbing his face, with two white slippers on the floor. The transcript panel (label 1) reads: “0 min 8 sec to 0 min 12 sec. The man sits on the edge of the bed, rubbing his face. Two white slippers sit on the floor.” In the conversation panel, the user asks (label 3), “Anything special about the slippers?” and the agent replies in a yellow box, “Yes, it appears that the white slippers resemble smiling faces at 0:10.” A small thumbnail (label 3a) below shows a close-up of the slippers, which resemble smiling faces. Labels 1, 2, 3, and 3a identify the transcript, timestamped video, user–agent conversation, and supporting thumbnail.}
    \vspace{-3mm}
\end{figure}

Clara quickly goes through the generated script and her attention is drawn to the time gap ``0 min 8 sec to 0 min 12 sec,’’ where the model mentioned a pair of slippers on the floor (Figure \ref{fig:contextaware}-1). Curious about why the AI chose to highlight this particular detail, she decides to investigate further. She uses the hotkey (\hotkeyName{Ctrl + 4}) to jump her focus directly to the conversational agent, which is context-aware and keeps track of which timestamp she was working on. \\

Clara asks: \texttt{``Anything special about the slippers?’’} 

The agent responds: \texttt{``Yes, it appears that the white slippers resemble smiling faces at 0:10” (Figure \ref{fig:contextaware}-3).}  

To understand how they resemble faces, Clara asks for clarification: \texttt{``What is it about the slippers that makes them look like they're smiling? Is there a pattern or are the slippers deformed?”} 

The agent provides a detailed explanation: \texttt{``This occurs because the two small holes on the slippers resemble eyes, a fold of material looks like a nose, and the purple band across the front forms a subtle smile.”} \\

This detailed explanation gives Clara a clear mental image. She refines the AI-generated line in the script editor to read: \texttt{``The man sits on the edge of the bed, rubbing his face. Two white slippers, resembling faces, sit on the floor.”} With the description finalized, her next step is to ensure it fits the time slot.

To check the timing and flow, Clara plays the narration of that line of AD synchronized with the video using the \hotkeyName{Ctrl + O} hotkey. She finds the generated AD script line is slightly too long to fit comfortably within the four-second gap.  She then removes ``the edge of’’ from the line to make it more concise. She plays the narration again, and the revised description now fits nicely within the gap.

Later, she reaches a quick montage of face-like objects that she needs to describe efficiently. She notices this section because the gap-identification step has produced a cluster of short adjacent gaps between 0 min 16 sec and 0 min 24 sec in the script editor. As she moves her screen reader focus through these lines, ADCanvas reads each cue's start time, end time, and duration, so she can tell that there are several short windows in sequence rather than one long gap.

She then asks the agent, \texttt{``List the objects that look like faces in this video and tell me the time that these appear.''} The agent responds with a sequential list:
\begin{itemize}[leftmargin=0.2in]
    \item A bathroom sink faucet and knobs. [0 min 16 sec - 0 min 18 sec]
    \item The back pockets of a pair of pants on a bed. [0 min 18 sec - 0 min 20 sec]
    \item A smiley face in the foam of a coffee mug. [0 min 20 sec - 0 min 22 sec]
    \item Two sunny-side-up eggs in a frying pan, forming a face. [0 min 22 sec - 0 min 24 sec]
\end{itemize}

While this list provides the factual ``what,’’ Clara’s end goal was to provide her creative ``how.’’ Recognizing a unifying theme, she uses her creative judgment to synthesize the agent’s list into a more evocative and concise description for the fast-paced montage: \texttt{``Faces follow him. In the bathroom sink, his own trousers, his morning coffee, and even his fried eggs.”}

After drafting descriptions for all the gaps, Clara performs a final quality control pass. She uses the \hotkeyName{Ctrl + I} hotkey to toggle the full AD track on and plays the video from the beginning. She listens to the complete experience, making minor wording and timing adjustments directly in the editor to ensure consistency and polish. Once satisfied, Clara uses the \hotkeyName{Ctrl + 9} hotkey to download the AD file and she shares it with her stakeholder. This iterative workflow between the script editor and the context-aware AI agent allows Clara to remain focused on the creative aspects of writing and timing descriptions. By handling the visual information retrieval and scaffolding the script, \texttt{ADCanvas} enables her to efficiently produce a high-quality AD track, addressing many of the accessibility barriers present in traditional authoring tools, such as the reliance on sighted interpreters for visual information and the inaccessible, granular controls common in mouse-and-slider-based interfaces. 

\subsection{Features \& Design Rationale}
The features of ADCanvas are designed to work in concert to achieve our three design goals.
\subsubsection{Design Goals}
The design of \textbf{ADCanvas} was guided by the barriers that BLV AD creators face with traditional authoring tools, which are fundamentally designed for sighted users \cite{bennett2025made,jiang2022co}. To create a system that is not only accessible but reimagines the creative process, we established three core design goals: 

\begin{itemize}[leftmargin=0.2in]
    \item \textbf{Enable independent authoring from BLV creators.} Our primary goal is to remove the need for sighted assistants created by inaccessible interfaces. The system should provide BLV creators with accessible control over the AD process, from initial video reviewing to final script export, using typical screen reader and keyboard-based navigation.
    \item \textbf{Facilitate video understanding.} To create authentic and high-quality AD, a creator first needs to build a mental model of the video's content, pacing, and emotional tone. Our second goal is to enable the system to describe video segments and answer questions. The system should allow creators to query the video's visual content at multiple levels of abstraction, from high-level summaries down to granular, timestamped details.
    \item \textbf{Shift cognitive load from visual information seeking to creative synthesis.} A major challenge experienced by BLV creators is the high effort needed to gather and keep track of visual information, especially when performing other tasks. Our third goal is to use AI to remove this burden, enabling the human creator to focus on the higher-order, creative aspects of AD: narrative construction, word choice, and tonal consistency. The system should act as a collaborative partner, providing all of the information that the creator needs so that they can focus on filtering the most needed information and crafting a script. 
\end{itemize}

\subsubsection{Enable Independent Authoring from BLV Creators}

To support independent, non visual authoring, ADCanvas is built upon a foundation of accessible controls and a keyboard first workflow. The system provides standard media controls, including Play, Pause, Forward, and Rewind, all operable through a screen reader. A key non visual affordance is the system's \textbf{audio cued timestamp feedback}. Whenever the video is paused, the system announces the current timestamp (for example, "Paused at 14 seconds" or "Forward to 19 seconds"). This cue offers users a basic sense of temporal position without requiring a visual timeline. It is important to note that this audio feedback does not replicate the full informational richness of a visual playhead. Instead, the timestamp announcements serve as lightweight temporal anchors that help users track their location during navigation and scripting.

The authoring interface itself is designed to be fully accessible. At its core is a WebVTT-based\footnote[1]{WebVTT is a standard file format for displaying timed text tracks, such as subtitles, captions, and video descriptions~\cite{w3c_webvtt_2019}. It is a plain text file that uses special cues with timestamps to synchronize text with media content, and it supports styling and positioning options for the text~\cite{w3c_webvtt_2019}.} script editor where users can write, review, and modify AD scripts. We found that traditional timestamp formats (e.g., 00:00:08) are not fluently read by novice creators through pilot testing,   so we adopted a more legible format (xx min xx sec)~\cite{w3c_webvtt_2019}. This user-centered adaptation improves the usability of the core editing interface, while still affording creators full control to manually write or edit descriptive text and adjust timestamps.

We complemented this interface with a workflow optimized for \textbf{keyboard-first navigation}. Dedicated hotkeys allow for not only media control (e.g., \hotkeyName{Ctrl + 1} for Play/Pause, \hotkeyName{Ctrl + 2} for Rewind 5 seconds, and \hotkeyName{Ctrl + 3} for Fast Forward 5 seconds), but also for rapid switching of focus between the script editor (\hotkeyName{Ctrl + 5}) and the conversational agent (\hotkeyName{Ctrl + 4}). These hotkey-based controls were designed based on pilot testing of user preferences and carefully mapped to avoid conflicts with common system-level and screen reader shortcuts across major operating systems (Appendix. Table \ref{tab:hotkeys}).  

To further reduce navigational effort and maintain the user's focus, the system implements \textbf{state persistence}. If a user navigates from the script editor to the agent to ask a question, their focus is brought back to the exact line they were previously editing upon their return. This persistence is important for maintaining mental context and reducing the cognitive load of re-orienting within the script. 

Finally, ADCanvas supports directly \textbf{exporting the finalized video} with the AD track and the original soundtrack mixed together, ready for sharing without further post-processing. 

\subsubsection{Facilitate Video Understanding through Conversation}
To help creators build a rich mental model of the video, ADCanvas transforms visual information retrieval into a natural, conversational dialogue. The system's \textbf{Visual Question Answering (VQA)} capability allows creators to probe the video's content at varying levels of abstraction. This enables an inquiry process that can move from high-level conceptual queries, such as ``Provide a brief summary of the entire video,'' down to specific, factual evidence like ``What visuals show the man is looking distressed?'' Users can ask about any visual element, from the color of an object to the emotional tone of an expression, allowing them to construct a detailed understanding of each scene. 

Beyond answering direct questions, the agent can proactively scaffold the creator's understanding of the video's structure and content. For example, the system can perform \textbf{gap identification} by analyzing the video's audio track to suggest silent periods where AD can be inserted, which draws an explicit connection between the entire video and the AD script. For example, few time gaps give a hint that the script will have to be concise and efficient. Furthermore, users can request \textbf{scene-specific descriptions} for particular timestamps. This provides a descriptive foundation for key moments, helping to build out the creator's mental model without requiring them to formulate a long series of granular questions.

\begin{figure}[t]
    \centering
    \includegraphics[width=1\columnwidth,alt={Screenshot of ADCanvas showing how the system updates references in the audio description script. On the left, the user asks, “Any chance the video mentioned the name of the man?” and the agent replies (a), “Actually, yes! The man’s name is Tom, according to the rolling credits at the end.” The user then requests, “Can you update the AD script for me accordingly?” and the agent responds (b), “Certainly! I’ve updated all 6 occurrences of ‘a man’ or ‘the man’ or ‘the guy’ to ‘Tom.’” On the right (c), the script panel displays tracked edits where red strikethrough text such as “A man,” “The man,” and “The guy” have been replaced with “Tom” in green, showing multiple corrections across the transcript. Labels a, b, and c mark the conversation and the updated script.}]{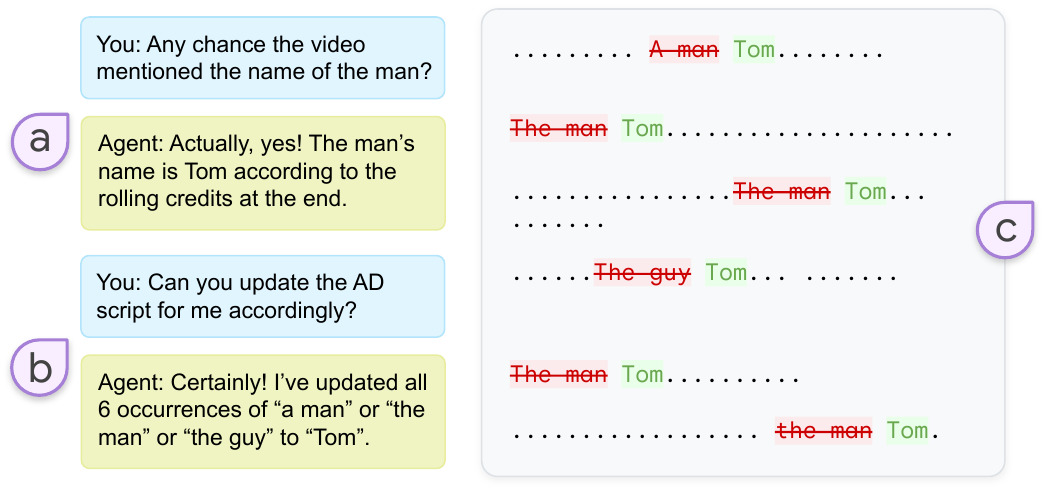}
    \caption{Global modification through the conversational agent. The participant wanted to know the man’s name (a), then decided to update the script with the man’s name by asking the conversational agent to update ``the man,” ``the guy,” and ``a man” to the man’s name (b). The agent made the global edit accordingly (c). Since it has a contextual understanding of the video/script, it did more than a string replacement. }
    \label{fig:global}
    \Description{Screenshot of ADCanvas showing how the system updates references in the audio description script. On the left, the user asks, “Any chance the video mentioned the name of the man?” and the agent replies (a), “Actually, yes! The man’s name is Tom, according to the rolling credits at the end.” The user then requests, “Can you update the AD script for me accordingly?” and the agent responds (b), “Certainly! I’ve updated all 6 occurrences of ‘a man’ or ‘the man’ or ‘the guy’ to ‘Tom.’” On the right (c), the script panel displays tracked edits where red strikethrough text such as “A man,” “The man,” and “The guy” have been replaced with “Tom” in green, showing multiple corrections across the transcript. Labels a, b, and c mark the conversation and the updated script.}
    \vspace{-3mm}
\end{figure}

\subsubsection{Shift Cognitive Load from Visual Information Seeking to Creative Synthesis}
To position the AI as a true creative partner, ADCanvas is designed to alleviate mechanical labor, thereby freeing the creator to focus on narrative artistry and creativity. This partnership is enabled by the agent's \textbf{context awareness}. To reduce the user's burden of repeating information, ADCanvas automatically includes the current video timestamp, the specific line of script being edited, the existing AD script, and the conversation history in its prompts to the AI (Appendix. \ref{prompt}). This allows the agent to interpret queries contextually, making the interaction feel like a fluid collaboration rather than a series of disconnected commands.

The most direct way the system offloads labor is through \textbf{full script generation}, where it automatically creates a complete AD script with descriptive text and timestamps through natural language from the users along with AD script~\cite{bittner2012audio,fryer2016introduction} and AD guidelines~\cite{AudioDes23:online,RehabilitationAct_Sec508}. This handles the often tedious first pass, allowing the creator to immediately engage in the higher-level work of refining and enhancing the narrative. To ensure user agency, ADCanvas also provides \textbf{flexible starting points}. Creators can choose their desired level of automation; other starting points include: drafting, where users can start with a script containing only pre-identified time gaps, allowing them to focus on writing the descriptions; or, from scratch, where users can begin with a completely blank slate, using the agent to ask visual questions about the video and manually create AD scripts independently.

Finally, this collaboration is supported by a tight, iterative workflow for generation and refinement. Creators can leverage the agent for \textbf{script editing} tasks, asking it to rephrase a line, shorten it, or provide alternative descriptions. The system supports both local and global updates, such as changing a character's name from ``a man'' to ``Tom'' across all script entries (Figure \ref{fig:global}).  Creators can then instantly verify the timing and flow of any change using the \textbf{line-specific narration preview} (through \hotkeyName{Ctrl + O}). This feature, powered by a state-of-the-art text-to-speech (TTS) model~\cite{comanici2025gemini}, plays the synthesized audio for a single line in sync with the video. Inspired by prior research \cite{natalie2024audio, Bragg2021expanding}, having adjusted AD script speed can enhance engagement and enjoyment of video experiences. The system automatically adjusts the narration speed based on the time gap and script length to ensure it fits, but maintains a 1x speed for longer gaps to preserve a natural cadence~\cite{Gemini2531:online}. For a holistic review, users can switch on the \textbf{global AD toggle} (through \hotkeyName{Ctrl + I}) and play the full AD track with the original sound. To clarify, only the AD narration is based on the TTS model, all the UI components, including conversational agent responses, buttons, and interacting with WebVTT AD scripts, are accessed through screen readers. This tight loop between AI-powered suggestions, direct user control, and immediate verification keeps the creator focused on the final creative output, enabling the kind of creative synthesis exemplified in the montage scene, where Clara transformed a factual list from the agent into the evocative narrative theme: ``Faces follow him.''

\subsection{Technical Implementation}
The ADCanvas prototype is a web application built with standard HTML, JavaScript, and CSS, ensuring broad accessibility and compatibility. We tested the interface across modern browsers, including Google Chrome, Apple Safari, and Mozilla Firefox. Crucially, We also ensured its compatibility with major screen readers, including JAWS~\cite{freedomscientific_jaws}, NVDA~\cite{NVAccess81:online}, and Apple's VoiceOver~\cite{Accessib64:online}. To enhance the stability of the authoring experience, we implemented client-side storage with IndexedDB~\cite{IndexedD81:online}. This ensured that all work-in-progress (both the script and the audio narration data) was retained on the user's device, protecting it from being lost due to inadvertent actions like closing a tab or refreshing the page.

The core generative features of ADCanvas are powered by the Gemini 2.5 series of multimodal models~\cite{team2023gemini}. Specifically, we leverage Gemini 2.5 Pro for the initial generation of AD, benefiting from its strong reasoning and creative capabilities~\cite{team2023gemini}. To power the interactive conversational agent, we use Gemini 2.5 Flash to ensure low-latency, responsive interactions~\cite{team2023gemini}. To balance accurate descriptions with creative nuance, the model temperature for all prompts to the AI model was set to 0.3. Additionally, the play line narration feature (activated through the \hotkeyName{Ctrl + O} hotkey) is powered by low-latency Gemini 2.5 Flash text-to-speech (TTS) model \cite{team2023gemini}. 
However, it is important to note that our primary contributions lie in the concept of AI-based AD co-creation by BLV users themselves and the design of the user interface and experience that enables this workflow. These contributions are independent of any specific model usage. We anticipate that these designs and interaction patterns will remain relevant and valuable as the underlying generative AI models continue to advance.

\begin{table*}[t]
\centering
\Large
\resizebox{\textwidth}{!}{%
\begin{tabular}{@{}llllp{21mm}p{35mm}p{108mm}@{}}
\toprule
\textbf{P\_ID} & \textbf{Age} & \textbf{Gender} & \textbf{Screen Reader} & \textbf{Visual \newline Condition} & \textbf{Professional \newline AD Experience} & \textbf{Content Creator Experience} \\ \midrule
1 & 45-54 & Man & VoiceOver & Legally Blind & Writing, \textgreater{}5 yrs & N/A \\
2 & 25-34 & Man & NVDA & Totally Blind &  Writing, QC, 1-2 yrs & Creates accessibility tutorials \& demos. (1-2 yrs) \\
3 & 25-34 & Man & NVDA & Legally Blind & Writing, QC, 3-5 yrs & N/A \\
4 & 55-64 & Woman & NVDA & Totally Blind & N/A & Creates YouTube videos for other blind people. (\textgreater{}5 yrs) \\
5 & 25-34 & Man & NVDA & Totally Blind &  QC, Narration, 3-5 yrs & Creates videos on blind game accessibility, tech, and lifestyle. (\textgreater{}5 yrs) \\
6 & 35-44 & Woman & JAWS, NVDA & Low Vision &  QC, 3-5 yrs & Manages social media for a small business. (1-2 yrs) \\
7 & 18-24 & Woman & JAWS & Totally Blind &  N/A & Creates content on accessibility, AI, and inclusive design. (3-5 yrs) \\
8 & 45-55 & Man & NVDA & Totally Blind &  QC, \textless{}1 yr & Manages social media for a business. (3-5 yrs) \\
9 & 35-44 & Man & NVDA & Totally Blind &  Writing, QC, \textgreater{}5 yrs & N/A \\
10 & 35-44 & Woman & JAWS & Legally Blind &  N/A & Creates educational videos on TikTok. (1-2 yrs) \\
11 & 18-24 & Woman & JAWS & Legally Blind &  N/A & Creates videos about her lifestyle as a blind woman. (\textless{}1 yr) \\
12 & 25-34 & Woman & JAWS & Totally Blind &  N/A & Creates product unboxings, accessibility videos. (3-5 yrs) \\ \bottomrule
\end{tabular}%
}
\caption{Demographic information of 12 BLV AD creators. The table includes each participant’s P\_ID, age, gender, types of screen reader they use, visual condition, experience with AD, and content creator experience. Note that all participants have over 5 years experience using AD, N/A means no professional AD or content creator experience.}
\label{tab:participants}
\vspace{-5mm}
\end{table*}

\section{User Study with BLV AD Creators}
To evaluate the utility of ADCanvas and investigate our research questions, we conducted a study with BLV AD creators and leveraged ADCanvas as a technology probe~\cite{hutchinson2003technology}. Our goal was to understand how a multimodal conversational agent and an accessible editor can support the workflows of BLV AD creators, identify the preferences of this human-AI co-authoring model, and derive insights to inform the design of future accessible authoring tools.

\subsection{Participants}
We recruited 12 participants (P1-P12) who self-identified as blind or low vision through community mailing lists, and professional networks. All participants were daily screen reader users (JAWS, NVDA, or VoiceOver) (Table~\ref{tab:participants}). Participants had to 1) be 18 years or above, 2) have vision impairments, and 3) have experience with AD creation or content creation (e.g., producing videos for social media). To ensure our findings would be relevant to a spectrum of potential users, we recruited individuals with varying levels of prior experience in AD creation. Participants self-reported how long they have used (not created) AD, whether they are AD creators , and if so, they self-reported their professional expertise (e.g., writer) and years of experience (Table~\ref{tab:participants}). All participants had over 5 years experience using AD. We also included their self-reported content creation experience (Table~\ref{tab:participants}). The study followed our organization’s research ethics approval process.

\subsection{Apparatus}
We conducted the study remotely on a video conference platform. Participants used their own computers, web browsers, and preferred screen readers, contributing to an ecologically valid environment. Researchers watched the session and listened to the screen reader's audio. Participants enabled their screen reader's speech viewer, which displays the text being read aloud, to further help the moderator follow along. We hosted the ADCanvas prototype on a private web server, which participants accessed via a unique link.

\subsection{Tasks and Materials}
We tasked participants with using ADCanvas to work on AD for three short video clips (Instructional (V1), Entertainment (V2), Documentary (V3)) each chosen to represent a distinct stage and challenge in the AD workflow. All three videos were intentionally selected to have minimal dialogue and clear natural pauses for audio description, allowing participants to focus on the authoring workflow rather than dense dialogue management. Each task presented a different starting point to evaluate the system's flexibility.

\begin{itemize}[leftmargin=0.2in]
    \item \textbf{Task 1: QC and Refinement.} For this task, participants worked with a cooking instructional video (\textit{3 Ingredient Nutella Mug Cake 2 Ways\footnote[1]{https://www.youtube.com/watch?v=sItYaC1z\_d0}} (V1) (02:00)). ADCanvas provided a fully AI-generated AD script, including both timestamps and descriptions. The goal was to assess how participants use ADCanvas to perform quality control (QC), by reviewing and editing an existing script.
    \item \textbf{Task 2: Authoring with Time Gaps.} Participants watched a short entertainment film (\textit{PAREIDOLIA - 1 Minute Short Film\footnote[2]{https://www.youtube.com/watch?v=JLmOkEEC9SQ}} (V2) (01:00)). For this video, ADCanvas identified the available time gaps for AD but provided no descriptive text. The task was to author a new AD script from this scaffold, evaluating how the conversational agent supports the initial drafting and creative process.
    \item \textbf{Task 3: Free-Form Authoring.} The final task used a clip from a nature documentary (\textit{Climate Change Impacts All of Us | National Geographic\footnote[3]{https://www.youtube.com/watch?v=QwLyscT3NgI}} (V3) (01:12)). Participants received a blank slate---no pre-generated script or time gaps. We designed this task to observe how users would approach AD creation from scratch.\looseness=-1
\end{itemize}

\subsection{Procedure}
Each study session lasted 120 minutes and followed a structured, four-part protocol:
\begin{enumerate}[leftmargin=0.2in]
    \item \textbf{Introduction and Setup (20 mins):} The researcher explained the study's purpose, obtained informed consent for audio and screen recording, and guided the participant through the technical setup.

    \item \textbf{System Orientation (10 mins):} Using a short explainer video, we gave participants a tutorial of ADCanvas. The researcher walked them through each of the core components: the video player, playback controls, script editor, and the conversational agent. Participants practiced using the essential hotkeys for navigation and media control and performed a test query with the AI agent.

    \item \textbf{AD Authoring Tasks (70 mins):} Participants then worked sequentially through the three main tasks (QC, Authoring from a Scaffold, and Free-Form Authoring)~\cite{Describi84:online}. This sequence represents a scaffolded progression of task complexity, allowing participants to focus on an evaluative task (QC) before advancing to guided (Scaffolded) and finally unguided (Free-Form) creative work. We asked them to use the think-aloud protocol \cite{charters2003use}, verbalizing their thoughts, intentions, and reactions as they interacted with the system. The researcher's role was primarily observational, intervening only for technical assistance. After each task, we asked a few follow-up questions about their experience, and they exported a log file of their interactions.

    \item \textbf{Post-Session Debriefing Interview (20 mins):} The session concluded with a semi-structured interview to gather qualitative feedback. Questions focused on the participant's overall impressions of ADCanvas, likes and dislikes related to each task, and the perceived quality and usefulness of the conversational agent's responses. Questions also inquired into participants’ broader perspectives on the potential role of AI in AD authoring and what features could improve the system. As part of this, we asked participants to rate the system's overall usefulness and their likelihood of using such a tool on a 7-point Likert scale~\cite{grier2013system,froehlich2025streetviewai}.
\end{enumerate}

\subsection{Data Collection and Analysis}
Our data consisted of audio and video recordings (including the screen share), text transcripts of the study sessions, and interaction logs that included the participants' input and system output. We conducted a thematic analysis of the data~\cite{braun2006using}. Three researchers first independently open-coded two study sessions each. We then synthesized the codes for these six study sessions into a codebook, and one researcher then applied the codebook to the remaining six study sessions. After coding, the first author then used affinity diagramming~\cite{lucero2015using} to develop the themes and sub-themes presented in this paper.

\renewcommand{\arraystretch}{1.35}

\begin{table*}[t]
\centering
\renewcommand{\arraystretch}{1.2}
\resizebox{1\textwidth}{!}{%
\setlength{\tabcolsep}{6pt}
\begin{tabular}{p{30mm} p{50mm} p{98mm}}
\toprule
\textbf{Interaction Type} & \textbf{Definition} & \textbf{Example Agent Commands} \\
\midrule
\textbf{Summarize Content} & Ask the conversational agent to summarize the video content into text description. & \texttt{``Summarize this video''} (P1) \newline \texttt{``Please can you give me an enhanced detail of the video summary''} (P3) \\
\addlinespace
\textbf{Generate Content} & Ask the conversational agent to generate timestamps of AD gaps, AD scripts, or both. & \texttt{``Generate descriptions for this video with time stamps.''} (P2) \newline \texttt{``Create a full description of the scenes in this video''} (P4) \\
\addlinespace
\textbf{Describe a Specific Scene or Object} & Ask the conversational agent to describe visual information of a scene or around an object. & \texttt{``Describe the scene in as much detail as possible.''} (P2) \newline \texttt{``Describe the white slippers in more detail.''} (P6) \\
\addlinespace
\textbf{Ask Visual Properties or Events} & Ask the conversational agent to provide additional visual information about certain visual properties or events. & \texttt{``What time does the watch appear?''} (P1) \newline \texttt{``Is there any conversational content in this video''} (P4) \\
\addlinespace
\textbf{Ask for Definitions or General Information} & Ask the conversational agent for definitions by leveraging the model’s world knowledge. & \texttt{``In the very beginning of the video, you said that two walruses were lying on ice floes. What is an ice floe?''} (P7) \\
\addlinespace
\textbf{Compare or Relate Information} & Ask the conversational agent to compare AI responses or different points in the video. & \texttt{``Can you describe the batter's texture now? How different is it from the two lines above, before the flour was mixed into it?''} (P6) \newline \texttt{``Describe this scene and its relation to the previous scene.''} (P8) \\
\addlinespace
\textbf{Revise Script} & Ask the conversational agent to revise the AD script directly. & \texttt{``Minimize ad text to comfortably fit into video, keep tone''} (P1) \newline \texttt{``Condense this wordy description into something that will fill a two second gap.''} (P6) \\
\addlinespace
\textbf{Adapt AI Responses to AD Scripts} & Ask the conversational agent to move responses from conversation to the AD script. & \texttt{``Incorporate the responses so far into the very beginning time stamps my focus is currently on.''} (P6) \\
\addlinespace
\textbf{Set Output \newline Constraints} & Ask the conversational agent to provide output with certain constraints. & \texttt{``Give a full and complete summary of this video using descriptive language. Do not editorialize''} (P8) \newline \texttt{``Did you make absolutely sure that the script does not step on any dialog or soundtrack or any production sounds within the video?''} (P9) \\
\bottomrule
\end{tabular}
}
\caption{Types of commands given to the conversational agent by our participants.}
\label{tab:typeofcommand}
\vspace{-5mm}
\end{table*}

\section{Findings}
In this section, we share the findings of our study in which BLV AD creators used ADCanvas for authoring AD scripts. This section is organized into 1) perceived usefulness of ADCanvas, 2) practices and creative workflows, 3) challenges and opportunities of AD co-creation with AI, and 4) reflections on AI as a co-author for AD.

\subsection{Perceived Usefulness of ADCanvas}
Throughout the study, participants shared feedback regarding aspects of ADCanvas’s usability that supported their workflow. They emphasized that their ability to maintain momentum hinged on the fluid navigation between AI interaction and script editing. The design of ADCanvas enabled users to move swiftly between writing, prompting, and reviewing.

Beyond observational insights, we asked participants to quantify the system's utility. When asked \textit{``how useful is ADCanvas’’} (on a scale of 1 to 7, where 7 is very useful, 1 as not at all useful), 10 out of 12 participants rated it a 7, and the remaining 2 rated it a 6. Similarly, when asked \textit{``how likely are you to use a tool like ADCanvas’’} (on a scale of 1 to 7, where 7 is very likely, 1 as not at all likely), 11 participants gave the highest rating of 7, with one participant giving a 6.

These scores offer strong evidence that participants saw meaningful value in the system—a tool they would seriously consider integrating into their creative workflows. For example, P1, who is an AD professional and content creator, said that \textit{``nothing else like this, now I can do everything.''} These high ratings further reinforce our qualitative findings, which reveal a system that participants viewed as a responsive partner in the creative process.

\subsection{Practices and Creative Workflows}
In this section, we show the practices and creative workflows that our participants engaged when interacting with embedded multimodal agents on AD creation. While we summarize tasks we witnessed in a prototypical order, participants frequently repeated actions (e.g., replaying the video), participants: 1) played the video \& generated a summary, 2) asked for visual details, 3) drafted a script, 4) verified and refined the script. 

\subsubsection{Playing the Video and Generating a Summary}

As a starting point, our participants often began by playing the video content, noting meaningful audio cues (e.g., dialogue, environmental sound) (Table \ref{tab:typeofcommand}). For example, P1 cued into a sound to predict what the main character was doing: \textit{``I can tell that the sound of water indicates that this is where he's washing his face.”}

After identifying the audio cues from the video, participants usually ask the conversational agent for a high-level summary (Table \ref{tab:typeofcommand})  (e.g., \texttt{``Summarize this video”} (P1), \texttt{``Generate a summary of this video for me”} (P11)). While they prompted for a ``summary,” they preferred they be verbose, as P2 commented: \textit{``Getting as much detail as possible and then filtering out what’s not important.”} Participants used the summary as a starting point and continued with multi-turn dialogue to understand more details of each scene.

\begin{table*}[t]
\centering
\renewcommand{\arraystretch}{1.2}
\resizebox{0.97\textwidth}{!}{%
\setlength{\tabcolsep}{6pt}
\begin{tabular}{p{30mm}|p{38mm} p{99mm}}
\toprule
\textbf{Goal of Inquiry} & \textbf{Information Category} & \textbf{Example Agent Commands} \\
\midrule

% ---- Visual Object Properties ----
\multirow[t]{4}{30mm}{\textbf{Visual Object Properties}}
& Color & \texttt{``What color is the watch?''} (P1) \\
\cline{2-3}
& Text, Symbols, and Logos & \texttt{``Does the mug have anything written on it?''} (P4) \newline
\texttt{``Describe the logo.''} (P8) \\
\cline{2-3}
& Texture and Material & \texttt{``Describe its [cake] texture.''} (P6) \\
\cline{2-3}
& Shape and Form & \texttt{``Describe the facial features of slippers.''} (P1) \newline
\texttt{``What size is the bowl?''} (P12) \\
\midrule

% ---- Events and Timing ----
\multirow[t]{3}{30mm}{\textbf{Events and Timing}}
& Character and Object Actions & \texttt{``Is there a hand holding spoon?''} (P1) \newline
\texttt{``What is the man doing?''} (P2) \\
\cline{2-3}
& Pace and Speed of Action & \texttt{``How fast is the mixture in the bowl being mixed?''} (P11) \\
\cline{2-3}
& Current State & \texttt{``So the cake has not been baked yet.''} (P5) \\
\midrule

% ---- Inference and Interpretation ----
\multirow[t]{2}{30mm}{\textbf{Inference and Interpretation}}
& Emotion & \texttt{``Why does this person look depressed?''} (P10) \\
\cline{2-3}
& Causality and Rationale & \texttt{``Why is a spoonful of mini chocolate added to the batter?''} (P3) \newline
\texttt{``What makes them [chocolate chips] optional?''} (P8) \\
\bottomrule
\end{tabular}
}
\caption{VQA information categories underpinning participant questions.}
\label{tab:qa_categories}
\vspace{-5mm}
\end{table*}

\subsubsection{Asking about Visual Details}
\label{visual details}
After getting a high-level overview of the video, our participants frequently used multi-turn dialogue to progressively build their understanding. They began with \textbf{scene-level questions} to grasp the video’s progression. For instance, after identifying a key sound, P4 requested a comprehensive description of the corresponding scene to establish a baseline understanding (e.g., \texttt{``Please create a full description of the scene with water sound in this video.”} (P4) (Table \ref{tab:typeofcommand})). This detailed overview then served as a foundation for more targeted follow-up questions.

As the dialogue evolved, our participants’ queries became increasingly granular, reflecting a layered process of visual inquiry grounded in their goals for accuracy, tone, and authorial control. They posed specific questions about \textbf{individual objects}, such as utensils, ingredients, and props, and sought to clarify visual properties such as color, texture, on-screen text, and shape (Table \ref{tab:qa_categories}).  For instance, P4 asked, \texttt{``Does the mug have anything written on it?”} and followed with spatial clarifications like, \texttt{``Where is the egg being cracked? On the side of the bowl or in the cup or on the counter?”}—probing not only object identity but its relationship to the surrounding environment.

In parallel, participants engaged the agent to understand dynamic elements, asking about \textbf{actions, timing, and pacing} (Table \ref{tab:typeofcommand}).  For example, P1 asked, \texttt{``Is there a hand holding a spoon?”}—a question seeking both confirmation of presence and potential alignment with narration. Similarly, P11 probed temporal rhythm: \texttt{``How fast is the mixture in the bowl being mixed?”} With these answers, our participants could choose which details to include in the script, when time allowed.

Finally, participants moved beyond surface-level description to request \textbf{interpretations} (Table \ref{tab:typeofcommand}), often grounded in affect, causality, or social cues. For example, P12 asked, \texttt{``How does the man appear distressed?”} prompting the agent to generate a narrative synthesis: \texttt{``The man appears distressed as he wakes up with a pained expression, wincing and bringing his hands to cover his face, as if in discomfort from the alarm.”} Understanding these interpretations supported AD scripting to align with the tone of the video.

\subsubsection{Drafting a Script}
ADCanvas enables multiple script-generation workflows, all of which our participants used during the study–auto-generating an entire script, inserting time gaps, and offering an empty edit field (Table \ref{tab:typeofcommand}).  To generate the full AD script, our participants used commands such as \texttt{``Please generate time gaps and audio description of what happens in this video scene by scene”} (P11) and \texttt{``Generate descriptions for this video with time stamps”} (P2). These commands yielded a full script which was annotated with time gaps where each line would be inserted, often including environmental settings, object-level details, and basic character actions aligned with the AD guidelines~\cite{AudioDes23:online,RehabilitationAct_Sec508}. Participants characterized the \textbf{auto-generated} scripts useful to build on. This approach minimized the friction of starting from scratch and was particularly valuable for BLV users who could not visually skim the video. As P11 emphasized, \textit{“Even if I don’t like everything it says, it’s helpful just to get it down and then adjust.”}\looseness=-1

When not auto-generating the full AD script, participants \textbf{authored from scratch} while using the conversational agent as an on-demand visual assistant. To do so, our participants manually composed each AD line, periodically querying the AI to fill in missing visual details, clarify timing, or verify actions (Table \ref{tab:typeofcommand}).  

Specifically, to obtain more information, several participants entered repetitive commands to learn more details about the video from the responses of the conversational agent (P2, P6, P8, P11). For example, P2 repeated the command, ``Describe this scene in as much detail as possible” about the same scene to try to obtain additional information, as exemplified by the following responses:\\

P2: \texttt{``Describe the scene in as much detail as possible.''}

Agent: \texttt{``The current scene at 3 seconds shows a close-up of a man in bed, from a high angle looking down. He has curly dark hair and stubble, and is wearing a white t-shirt. His eyes are closed, and his expression is a sorrowful grimace…”}

P2: \texttt{``Describe the scene in as much detail as possible.”}

Agent: \texttt{``At the 3-second mark, the scene presents an extremely close-up, high-angle view of a man lying in bed. He possesses thick, dark curly hair and noticeable stubble on his face. Dressed in a simple white t-shirt, his eyes are closed, and his facial expression is a clear sorrowful grimace, with his brow furrowed and mouth slightly downturned…''}\\

P6 used an inverted process, by writing lines first and asking questions after: \textit{``I like doing it myself first. Then I check if the AI saw something I didn’t.”} This practice preserved P6’s voice while engaging the AI as an auxiliary set of eyes.

Taking another approach, some participants requested the conversation agent to make specific edits. To accomplish this, some participants scripted more manually, by copying the response (Figure \ref{fig:teaser}-c)  and pasting it in the AD script editor for refining. Like P5, they also asked the conversational agent to refine lines according to parameters: \texttt{``Can you edit this line to include the on-screen instructions for cooking?”} (P5) (Figure \ref{fig:teaser}-d). While these participants leaned into the conversational agent to draft entire lines of AD, they still directed the process granularly to maintain agency over its development.

\subsubsection{Verifying and Refining the AD Script}

After initializing the scripts, our participants also used the agent to revise and verify script content. Upon writing or receiving a draft line, they  incorporated content from that line into confirmation and verification questions (Table \ref{tab:typeofcommand}),  such as:\\

% \texttt{``How was the cake baked?’’} while working on \texttt{``The finished cake has puffed up over the mug's rim.’’} [P5]

% \texttt{``Is there a microwave in the video during this timecode?’’} while working on \texttt{``The finished cake has puffed up over the mug's rim.’’} [P5]

\texttt{``Is there a hand holding spoon?’’} while working on \texttt{``A spoon scoops a bite from the gooey cake.’’} (P1)

\texttt{``Why does it say that things look like faces? describe the facial features of slippers’’} while working on \texttt{``He sits up, showing face-like slippers.’’} (P1)

\texttt{``Where is the man looking?’’} while working on \texttt{``Waking up, a white man slaps his alarm clock.’’} (P9)\\

This interaction allowed participants to verify visual details to the conversational agent to preserve their own narrative authority. As P2 explained, \textit{``I like to verify some of the information and then have follow-up. It’s a lot of back-and-forth.”} 
Participants appropriated VQA for verification (Table \ref{tab:typeofcommand}),  where they interpreted additional contextualized detail in answers as a signal that AI-generated AD lines were appropriate. For example, while working on V1’s AD script, P5 asked questions about the baking process since there were few details in the auto-generated AD script and P5 found the cake was ``finished” suddenly \texttt{``[0 min 41 sec to 0 min 44 sec] The finished cake has puffed up over the mug's rim”}:\\

P5: \texttt{``How was the cake baked?”}

Agent: \texttt{``Both mug cakes featured in the video were baked in a microwave for 50-60 seconds at 1000W.”}\\ 

He then questioned whether the microwave mentioned in an AI response was in the video since it was not mentioned in the auto-generated AD script:\\

P5: \texttt{``Is there a microwave in the video during this timecode?”}

Agent: \texttt{``The microwave itself is not shown in the video at the 36-second mark. However, immediately after this, from 39 to 42 seconds, text appears on screen indicating that the cake will be cooked in a microwave for 50-60 seconds.”} \\

Realizing the only reference to a microwave was in text, P5 asked the conversational agent to include the screen instruction to the AD script to make it clear regarding the source of the information:\\

P5: \texttt{``Can you edit this line to include the on-screen instructions for cooking?”}

Agent: \texttt{``I've updated the audio description to include the on-screen cooking instructions. I've split the original line at 36 seconds into three shorter segments to accurately reflect the changes in the video and accommodate the new information. The script now includes details about microwaving the cake.”}\\

This iterative verification process shows how our participants leveraged the conversational agent to understand and verify the visual content from the auto-generated AD scripts, which supports the accessible quality control process by BLV AD professionals without sighted interpreters.

\subsubsection{Model Errors}
Overall, the multimodal AI model generally provided accurate information, although as noted elsewhere in the paper, the specifics of the model’s responses and actions did not always match user expectation. To give a sense of overall accuracy of the model itself, two of the authors reviewed the 203 prompts entered by study participants and 202 system responses (one prompt was repeated due to a technical issue) and coded them independently, then discussed and resolved disagreements. We counted two types of inaccuracy: incongruent model responses and VQA errors. These made up 22 of the 202 responses (10.9\%). Incongruent model responses occurred when the model took an action or gave a response that did not match the intent of the participant’s prompt; there were 18 such events (13.4\% of 134 responses), the majority involving the system taking an action beyond what was asked, such as editing the AD script in response to a question about video content. VQA errors occurred when a response to a VQA prompt contained some false information; there were 4 such events (5.9\% of 68 VQA responses). While this does not provide a complete picture of the model’s accuracy, for example this does not address the comprehensiveness or level of detail of scene descriptions, it demonstrates that current models are able to provide high accuracy in interpreting user input and answering visual questions.

\subsection{Challenges and Opportunities of AD Co-Creation with AI}
In this section, we highlight design challenges and opportunities suggested by our participants after using ADCanvas.

\subsubsection{Separating Agent Conversation from Editing}
A primary source of confusion was the AI's tendency to be proactive, often taking actions that users did not explicitly request. The most common instance was the AI automatically inserting its generated text into the AD script after being asked a purely informational question. This occurred because the model interpreted certain questions as cues to generate or revise AD, so the system responded by inserting or modifying script lines even when users intended only to ask for information. This unsolicited action, while sometimes helpful, repeatedly violated the user's sense of control and authorial agency. P12 responded to ADCanvas’s unsolicited  edits: \textit{``I was a little surprised that Gemini puts stuff into the editor. Like I didn't expect that.''} P8 reacted with more alarm when the AI updated his script without permission, exclaiming, \textit{``Whoa...I did not tell you to do that''} (P8).

This behavior highlights a fundamental design tension: ADCanvas’s attempt to be helpful by anticipating user needs sometimes overstepped its role as an assistant. Participants expressed a clear desire for the final say on what gets added to their creative work. As P12 suggested, \textit{``it might be nice to have it wait for uh the go-ahead to add stuff, because, you know, we might be asking for information but then... decide that it isn't that relevant after all.''} This led to specific design recommendations centered on explicit user confirmation. P8 proposed a simple toggle: \textit{maybe it's a checkbox or maybe you can lock and unlock or something.''} P6 envisioned a collaborative model where the AI's suggestions would not overwrite their own work: \textit{``I think if it could be put like another line um under what I wrote...yeah, that's an important distinction is just to leave that [P6’s AD script] alone.''} Participants also suggested the system provide visual or auditory cues before executing any insertions, which would let users assess and confirm suggested actions without disrupting their workflow (P5, P8, P10, P12).

\subsubsection{Supporting Variable Levels of Timing Precision and Audio Control}

Beyond AI interaction, our participants had different preferences for granular or simplistic controls, related to their level of expertise~\cite{cheema2025describepro}. Regarding granular and precise control, some AD professional participants (P1, P5) indicated that there is a need to go beyond basic functionality and offer the kind of granular power found in professional creative software. For example, P1 commented: \textit{``I want to be able to adjust the timestamps [in the AD script] in milliseconds.”} These requests focused on giving the user finer-grained control over the audio mix, timing, structure, and navigation of the AD script. To improve the time control, our participants also requested more granular playback controls, such as the ability to \textit{``skip by a second or even skip by frames''} (P5). This was complemented by the need for a hotkey that would simply report the current playhead position, as P1 noted they were often unsure \textit{``exactly where you are in the timeline''} (P1). In contrast, some of our participants who were not AD professionals (P4, P11) preferred an uncomplicated interface: \textit{``I like how this system is simple enough to make it accessible for me to use”} (P11). Therefore, this shows a need to offer different settings for AD creators with different levels of expertise.\looseness=-1

\subsubsection{Matching AD Narration with Video Content}
Especially valued for its role in quality control was ADCanvas’s generation and playback of text-to-speech AD output, at either the AD line (\hotkeyName{Ctrl + O}), or entire script (\hotkeyName{Ctrl + I}) level. P5 commented: \textit{``I thought it was cool to be able to, like, hear a snippet of what that specific line was. That's something I have always struggled with.”} P1 stated, \textit{``Sometimes I want to hear what it sounds like before I change it. This way I don’t guess,”} indicating how reviewing mixed AD enhanced their QC process, compared to scrutinizing text alone.

Participants were excited to use this feature but wanted even more control over speech output, such as being able to adjust the volume of each soundtrack (P11). While audio ducking is standard in mixing the final AD with the soundtrack, P1 noted its utility for QC: \textit{``So you can mix like 80\% [AD] audio, 20\% [video] audio, not to say you're going to render it out that way, but just maybe for monitoring.'} P1 also recommended that users be able to test different gendered and toned narrations, sharing: \textit{``the [AD] voice [for V3] then sounded too happy and that's why I was going to ask if it could change the voice.”} This corresponds to prior work on AD customization from the AD consumer perspective~\cite{cheema2025describe, natalie2024audio}.

\subsubsection{Making Common Tasks Easy and Uncommon Tasks Possible}  
ADCanvas features hotkeys for some common actions, such as switching between the conversational agent and script editor. Participants frequently used and praised these hotkeys—P11 described these transitions as \textit{``helpful for not breaking my flow.”} P12 noted that hotkeys were useful because, \textit{``I didn’t feel like I had to stop what I was doing just to ask a question or fix something. I could jump right to the agent, ask it, and jump back in [to the AD script] where I left off,”} highlighting how minimal navigation keystrokes enhanced her AD creation.

As participants discovered their preferred set of commands, they sometimes suggested that these common commands receive their own hotkeys. The task of repeatedly typing commands like \textit{``describe this scene''} led multiple users to suggest a hotkey for \textit{describing the current scene}. P2 commented, \textit{``It might be helpful just to, uh, have, like, a hotkey that would kind of give you a description especially based on you know the frame that you're paused at''}. Taking this idea further, P4 suggested a system for storing commonly used prompts and being able to retrieve them later, \textit{``that would do the initial things that I want to do every time''}. Participants’ desire for the system to anticipate common tasks also appeared in requests for more proactive output from the AI, such as allowing the AI to watch a scene and proactively suggest potential descriptions or AD gaps, which the user could then confirm or reject. This model mirrors collaborative writing environments in which suggestions can be reviewed, accepted, or ignored.

\subsubsection{Customizing Output and Developing Consistent Styles} 
As participants became more familiar with the interface, and were able to generate AD content, they often started to consider the challenge of producing consistent AD content throughout a video or a single video. While these problems were not relevant for our short study tasks, participants noted the importance of maintaining a consistent style in real world use cases. One way to achieve this would be allowing the user to create a custom prompt for generating AD. P1 suggested that this might take the form of ``project notes” where the AD creator could include client instructions, project goals, key themes, or style guidance.

Another series of suggestions involved features to support users in learning how to write better AD, such as built-in educational resources. P9 wished for a \textit{cheat sheet} with useful prompts to guide him. P1 offered a more detailed recommendation for a guide on \textit{best practices}, suggesting the tool could offer an \textit{``index or something like that’’} with example prompts that users could \textit{``experiment with’’}. This would not only help novice users but would also empower experienced users to unlock the full potential of the AI partner.

\subsection{Reflections on AI as a Co-Author for Audio Description}
\subsubsection{Resolving Misalignments between User Expectations and AI Output}
Participants negotiated AI responses which misaligned with their expectations. Commonly, misalignments occurred when participants either wanted more or less objectivity. First, participants noted the AI often generated ``a lot of fluff'' (P6), by injecting interpretive language that had to be edited out to maintain the \textbf{objective tone} required by AD standards~\cite{AudioDes23:online,RehabilitationAct_Sec508}. P6 instructed the conversational agent to remove interpretations from a video summary:\\

P6: \texttt{``Please start by summarizing the entire video…”}

Agent: \texttt{``The video opens in the early morning with a man reluctantly shutting off his alarm to the loud sound of a buzzer, accompanied by the distant sounds of dogs barking and a rooster crowing. Initially groggy and sullen, his mood slowly brightens as he goes through his morning routine…”} 

P6: \texttt{``Please summarize the video without making assumptions about how the character is feeling.”}

Agent: \texttt{``The video begins with a man in bed as an alarm rings, accompanied by dog barks and a rooster crowing. He sits up and puts on white slippers designed to resemble faces…”} \\

The first summary P6 received strayed from describing literally what was available visually to speculating on the man’s emotions and motivations for certain actions. The second response, after P6’s instructions, focused more on the events in the video, providing a useful starting point for P6.

In other instances, the conversational agent \textbf{did not provide information} that participants needed to write AD. P9 considered ethnicity ``crucially important” to support representation. Yet the conversational agent relayed that such an assumption was out of bounds:\\

P9: \texttt{``What ethnicity is the man?”}

Agent: \texttt{``I cannot determine the man's ethnicity from the video. As an AI, I avoid making assumptions or categorizations based on appearance.”} 

P9: \texttt{``what is the skin tone of the man in bed?”}

Agent: \texttt{``The man in bed appears to have a light to medium skin tone.”} \\

P9 knew from experience that he would have to try ``prompt injection”--rephrasing the question to get a more favorable response. By asking about skin tone, he could begin to incorporate appearance information into his AD. However, his need for certain interpretations misaligned with the interpretations within the AI’s range of responses.

Finally, participants’ thresholds for objectivity not only concerned descriptors but their \textbf{presentation}. Reading an AI-generated AD line for V1, \texttt{``Optional chocolate chips are added and stirred into the batter,”} P8 asked, \textit{``Why does it say option? That seems like it's imposing.”} P8 was confused how ``optional” as a concept could be shown visually. He took it up with the conversational agent to get clarity:\\

P8: \texttt{``Where are they described as optional?”}

Agent: \texttt{``The chocolate chips are described as optional in line 43 of the AD script, from 1 minute 31 seconds to 1 minute 36 seconds.”} \\

Finding the script reference irrelevant when his question was about the video, it occurred to P8 that the information may be presented differently, asking:\\

P8: \texttt{``Is there text displayed?”}

Agent: \texttt{``Yes, at the current timestamp, the text `CHOCOLATE CHIPS OPTIONAL' is visible at the bottom left…”} \\

P8 pointed to subtle presentation differences which keep the AD objective: \textit{``I changed it [AD] to `text, optional,’ because otherwise the AD is inferring from the video. Even though it's accurate…it’s  the little nuance to how it's presented.”}

Participants had different thresholds regarding objectivity and interpretation, which the AI responses did not always align with. From removing ``fluff” (P6) to seeking information like P9’s interest in describing appearance which the AI considered too subjective to assume, participants had to negotiate these misalignments with follow-up questions and commands to craft AD that met their differing needs for objective and selectively interpreted AD. This concerned not only the terms included in the script, but how they were framed, such as clarifying text instructions.

\subsubsection{Preferred Roles for AI in AD Creation}
After the study tasks, we asked participants what role AI should play in AD authoring. To show their range of responses, we focus on their perspectives regarding using AI in \textbf{quality control (QC)}. Participants largely agreed that QC, ensuring AD conforms to standards and is understandable to BLV audiences, requires human oversight. P9, a QC professional, stressed that QC is ``a conversation” between reviewer and writer, giving an example of a QC-related question a writer might receive: \textit{``Hey, what made you decide on putting a cashmere sweater here rather than just a regular sweater?”} He explained the term might hint at class status—\textit{``Does the cashmere point to her class?''}—and should only be removed if that nuance was preserved elsewhere. P9 also warned that AI might erase collaborative voices in AD production.

Others saw AI as a helpful assistant for lower-level tasks. P6 described asking AI to confirm visual details, \texttt{``What color does this...character in a red shirt have?”} and then verifying them manually. P8 echoed this view, imagining AI assisting with spelling or checking narration fit without taking over interpretive judgments.

When discussing \textbf{tensions between efficiency and expertise}, content creators highlighted how tools like ADCanvas could improve both efficiency and accessibility. P10, who posts frequently across platforms, noted how real-time narration requires shifting perspective, e.g., ``your right side through the video”, and praised the time savings ADCanvas could offer. P12 envisioned using it for both professional and personal videos, even for family content: \textit{``Just getting information about the content that I’m receiving…would be really nice.”} P11 saw broader impact: creating \textit{``shared descriptive experiences”} so BLV people could \textit{``have a conversation about [videos] just like anybody else.”}

However, concerns remained. P5 feared that tools like ADCanvas might replace blind professionals, saying: \textit{`I would probably boycott using this in any professional setting,”} citing risks to both job security and AD quality. For P5 and others, responsible AI use meant centering the perspectives of BLV professionals and ensuring that those the AD serves are the ones leading its creation.

\section{Discussion}
Our findings demonstrated the practices of BLV AD creators using an MLLM-based AD co-authoring tool, characteristic of desired human-AI collaboration workflow, and design opportunities for accessible MLLM-based AD creation systems. In this section, we will discuss 1) reflection on independent AD creation by BLV users, 2) users’ need for customization, 3) tensions within AI-assisted AD authoring, and 4) multimodal interface design.

\subsection{Can BLV Authors Independently Create AD?} 
Our findings suggest that ADCanvas represents a meaningful step forward in enabling BLV creators to independently produce high-quality AD scripts. Participants consistently expressed that the system removed traditional dependencies on sighted collaborators by offering a fluid, screen-reader-accessible interface and a context-aware conversational agent. The conversational workflow empowered users to independently explore visual content, generate scripts, and iteratively refine descriptions, all through non-visual modalities. Participants acted as directors of the AD process: they exercised judgment, rejected inappropriate outputs, scaffolded their own script structures, and guided the AI with nuanced prompts. These interactions affirm that ADCanvas succeeded in supporting independent authoring in principle and in practice, with participants rating its usefulness and likelihood of future use at the highest levels.

On the other hand, our study also surfaced clear boundaries regarding what this independence currently entails. First, participants still engaged in a significant process of verification, often asking follow-up questions to clarify ambiguous or overly interpretive responses from the AI. In some cases, the agent failed to honor professional AD conventions (e.g., neutrality, precision, or descriptive hierarchy), prompting additional editing and oversight. Second, while users could control fine-grained elements such as timing and structure, the need to “negotiate” the AI's behavior revealed an asymmetry: rather than configuring the system upfront to follow a specific style or rule set, users were burdened with maintaining that alignment through repeated conversational scaffolding.

\subsection{Desire for Customization and Control}
Our study revealed that while ADCanvas supported multiple entry points into the authoring workflow, it did not fully meet participants’ needs for controlling how the AI contributed. Participants consistently expressed a desire to customize the agent’s behavior, such as requiring confirmation before applying changes or choosing when to receive suggestions versus direct edits \cite{ClaudeCo7:online}. Although the system enabled users to generate scripts, ask visual questions, or work manually, it lacked mechanisms to explicitly define the agent’s role or level of initiative. This led to moments of friction; for instance, some participants were caught off guard when ADCanvas modified script content without their consent (e.g., auto-updating lines after a conversation), which disrupted their editing flow and reduced their sense of control.

These findings highlight the importance of agent configurability as a design principle for accessible creative systems~\cite{zhang2023imageally,othman2023fostering}. Participants wanted to adapt the system’s behavior to match their personal authoring styles, whether that meant toggling between consultative and generative modes, or adjusting how much authority the AI had during different stages of the task. Some preferred a passive assistant that answered questions and offered optional suggestions, while others wanted more active support when starting from a blank slate. This aligns with work showing that customizable interaction models help users maintain creative rhythm and editorial authority in co-creative workflows~\cite{jiang2023beyond, petridis2024constitutionmaker}.

Going forward, accessible authoring tools should treat interaction mode selection (e.g., suggest-only, verification-required, automatic generation) as a core feature. By enabling users to define agent behavior and boundaries at the level of interaction, future systems can better support both consistency and autonomy in the creative process~\cite{liu2023wants, liu2025gensors}.

\subsection{Tensions within AI-Assisted AD Authoring}
\subsubsection{Trust vs. Verification}
While participants appreciated ADCanvas’s conversational affordances, they did not treat the AI as a flawless oracle. Instead, they adopted a “trust but verify” stance, using the agent’s suggestions as starting points, then rigorously validating them. This reflects prior work in human-AI collaboration, where trust depends not just on correctness but also transparency, explainability, and human oversight~\cite{gonzalez2024investigating, mozannar2023effective}. Rather than being discouraged by verification, users developed strategies to navigate it, similar to how sighted users tolerate “creative errors” in co-creative systems~\cite{jiang2024survey}. In professional AD, where objectivity, tone, and timing are critical, unverified AI suggestions risk undermining narrative quality and representational accuracy. Users often rejected overly interpretive or emotionally assumptive descriptions (Section \ref{visual details}), reinforcing the importance of a neutral narrative voice~\cite{snyder2008audio}. Still, some users wanted limited interpretation to match tone or convey character traits. Trust increased when the agent acknowledged ambiguity or responded conservatively to uncertain prompts. These findings highlight the need for contextualized responses, e.g., justifying interpretations with visual evidence or timestamps, and systems that surface uncertainty via confidence scores or multiple candidates~\cite{macleod2017understanding}. Allowing users to inspect reasoning, flag unreliable outputs, and fine-tune responses fosters trust. Exposing uncertainty (e.g., “I’m not sure” or “here are two options…”) helps users decide when to trust or verify.

\subsubsection{Simplicity vs. Professionalism}
Our participants, especially those with AD production experience, advocated for professional-grade precision, including millisecond-level timestamp control, adjustable audio mix levels, and control over narration pacing. These preferences contradict the notion that accessible interfaces should be ``simplified” for BLV users. Instead, participants expressed interest in tools that scale with expertise. Some users appreciated ADCanvas’s minimalist interface, while others sought advanced options akin to those in mainstream DAWs. This tension suggests a need for adaptive complexity: tools that can start simple but reveal deeper layers of control as user needs grow. 

This principle echoes prior findings in assistive technology and creative tooling, where users with disabilities are not merely consumers of simplified systems, but domain experts demanding high-fidelity control over their craft~\cite{zhang2025vizxpress,cha2024understanding}. We recommend future accessible authoring systems to provide toggleable “modes” (e.g., Basic, Intermediate, Professional), scripting APIs, and exportable control layers (e.g., XML/JSON).

\subsubsection{Assistance vs. Autonomy}
Our findings suggest that MLLM-powered tools like ADCanvas hold strong potential for both professional and personal use in AD creation, with all participants expressing interest in using its features for drafting and quality control. Participants valued the system’s ability to streamline tasks such as scene summarization and gap detection, aligning with prior work on AI-assisted creativity and productivity tools~\cite{demirel2024human,morris2023design}. However, several professional AD writers raised concerns that such automation could diminish the value of their labor or eventually displace their roles, reflecting broader tensions in creative industries around authorship, job security, and AI integration~\cite{fukumura2021worker}. Rather than full automation, participants favored collaborative, human-in-the-loop systems that preserved creative control, emphasizing the importance of transparency, customization, and respect for domain expertise~\cite{amershi2019guidelines}. Given the nuanced, narrative nature of AD, future design should prioritize co-creative support over replacement, ensuring that tools adapt to professional practices while enhancing accessibility outcomes.

\subsection{Reflection on the Multimodal Interface Design}
Our findings show that ADCanvas benefits from coordinating conversation, text editing, and audio playback in a unified workflow. At the same time, participants surfaced several focused interface improvements that would make ADCanvase more useful and useable. Below, we summarize the most concrete needs identified in the study.

\textbf{Separating conversation from editing and keeping suggestions controllable.}
Participants wanted to ask questions without risking unwanted changes, and they expressed mixed preferences for proactive output. A future design should provide two clear modes: an Ask Only mode where the agent never edits the script, and a Review Suggestions mode where potential edits appear in a separate pane and require explicit confirmation. To keep suggestions non-intrusive, ADCanvas could include a suggestion drawer that quietly collects agent-generated options for wording, timing, or gap placement. Users could skim, accept, or discard these suggestions with a single keystroke, supported by short auditory cues indicating when the agent is about to modify or suggest a line. This preserves creative control while still allowing users to benefit from the agent’s assistance.

\textbf{Reducing repetitive prompting.}
Many users reused the same queries, such as describe this scene or list objects at this timestamp. The interface could offer configurable prompt shortcuts that can be triggered by hotkeys and automatically reference the current timestamp or selected line. A small library of reusable prompts would reduce conversational overhead and speed common workflows.

\textbf{Providing accessible timing and audio controls.}
Participants valued hearing mixed TTS and video audio but wanted more precise control. A future version should expose simple, screen reader friendly controls for AD volume, original audio volume, ducking strength, and narration voice. Timing tools should allow both coarse and fine adjustments, including second level or sub second changes for expert users.

\textbf{Supporting different levels of authoring expertise.}
Some participants preferred a simple surface, while others wanted professional grade control. ADCanvas should offer adjustable complexity, such as a basic mode for high level drafting and an advanced mode that reveals detailed timing, editing, and mixing features. These layered controls would help the system adapt to creators with varying goals and experience.

Together, these refinements would reduce repetitive user actions, and make the multimodal interface more predictable and better support detailed creative control, while maintaining full non visual accessibility.

\section{Limitations and Future Work}
Our study, while rich in insight, is constrained by its sample size and lab-based setting. Participants worked with a fixed set of video materials, which may not represent the diversity of content encountered in real-world AD projects. For example, these projects could be longer, have lots of onscreen text, or be in different languages. A further limitation is that we tested ADCanvas using videos that did not include dialogue. We focused the study on the core authoring workflow rather than on audio interference management. While ADCanvas supports timestamp adjustments when dialogue is present, we did not evaluate this capability in a dialogue-rich setting. Future work should include videos with overlapping speech to examine how non visual tools can better support timing decisions and AD placement in more complex audio environments. Future work should additionally explore longer-term deployments in authentic production contexts and expand customization to support collaborative workflows. There is also an opportunity to investigate how these tools scale across genres (e.g., film, educational content) and production roles (e.g., writers, narrators, quality reviewers). 

Moreover, ADCanvas focuses on AD authoring and enables real-time Gemini TTS narration for QC and exporting embedded AD in the video. However, additional features could provide full control for narration, such as supporting human narration or changing narration characteristics~\cite{natalie2024audio}. ADCanvas also does not currently support waveform inspection, gap visualization, or fine-grained timing alignment. These timeline-based cues are central in professional AD editing and allow sighted editors to quickly assess gap length and structure. Our system provides only partial temporal awareness through timestamp cues and TTS previews, so we position ADCanvas as an exploratory non visual authoring approach rather than a full alternative to visual timeline editors. Furthermore, our evaluation did not benchmark model accuracy systematically; we report observed correctness from in-situ use. Future research should compare different models’ performance for multimodal AD authoring support.

\section{Conclusion}
In this work, we presented ADCanvas, an accessible authoring tool leveraging multimodal models and a conversation interface, which reimagines the process of AD creation for BLV creators. By shifting away from visual timelines toward a dialogic, principle-driven paradigm, ADCanvas empowers users to author, refine, and control AD scripts through natural language interaction and keyboard-based navigation. Our study with 12 BLV participants revealed that the system meaningfully supports a range of creative workflows, serving as an information conduit, a structural drafter, and a junior partner in the authoring process, while also surfacing important needs for trust calibration, authorial control, and professional-grade precision. Participants embraced the system not only as an accessible tool, but as a springboard for independence, productivity, and creative ownership. We contribute empirical findings that unpack the nuanced dynamics of human–AI co-creation in non-visual media production, and we offer concrete design implications for future accessible tools. 

As LLMs and multimodal agents continue to evolve, our work affirms the critical need to center user configurability, transparency, and agency, particularly for communities long excluded from traditional creative tools. ADCanvas charts a path toward a more inclusive future of media authorship, one where creative control is not constrained by interface modality, but is expanded through Human-AI collaboration.

\begin{acks}
We thank Philip Nelson for his support and constructive feedback. We sincerely thank our participants for their time and for sharing their experiences, which made this work possible. We also appreciate the reviewers for their thoughtful and helpful feedback.
\end{acks}


\bibliographystyle{ACM-Reference-Format}
\bibliography{sample-base}

\onecolumn
\appendix

\section{ADCanvas Hotkeys}
\label{Hotkeys}
\begin{table*}[h]
\centering
\renewcommand{\arraystretch}{1.3}
\begin{tabular}{p{32mm}|p{90mm}}
\toprule
\textbf{Hotkey Combination} & \textbf{Functionality of the hotkey} \\
\midrule
\texttt{Ctrl + 1} & Play/Pause Video \\
\texttt{Ctrl + 2} & Jump 5 seconds back \\
\texttt{Ctrl + 3} & Jump 5 seconds forward \\
\texttt{Ctrl + 4} & Move to conversation agent \\
\texttt{Ctrl + 5} & Move to text editing area \\
\texttt{Ctrl + O} & Play audio for current segment \\
\texttt{Ctrl + I} & Toggle AD Track \\
\texttt{Ctrl + M} & Regenerate the AD script \\
\texttt{Ctrl + 9} & Download all logs and export video with AD \\
\texttt{Ctrl + L} & Read out current line number \\
\bottomrule
\end{tabular}
\caption{Hotkeys for ADCanvas.}
\label{tab:hotkeys}
\end{table*}

\section{Prompt}
\label{prompt}

\begin{lstlisting}

Process this user's command: "{{command}}"

# Application
The user is interacting with an application that is being used to create audio description (AD) for blind or visually impaired viewers, for a short video. Below we include the user's command, the link to the video, the current timestamp, the current AD script, and the line that the user is on. Some commands may relate to the specific location in a video, such as asking about the content of the video frame, or asking for changes to a specific part of the AD script.
The script is in WebVTT format, so that each AD utterance will be on two lines, the first with timestamps and the second with the AD text. An (truncated) example would be something like:

===== START OF AD SCRIPT (NOT A LINE IN THE SCRIPT) =====
0 min 10 sec to 0 min 13 sec
The fox walks slowly through the snow.

1 min 21 sec to 1 min 25 sec
The fox approaches, looking around.

2 min 50 s to 2 min 55 sec
White fox sprints toward distant carcass.
===== END OF AD SCRIPT (NOT A LINE IN THE SCRIPT) =====

Note that the start and end timestamp must be in this format: "1 min 21 sec to 1 min 25 sec".
Also note that each "segment" contains two lines, the first line contains the start and end timestamp, the second line contains the text. Each segment is separated by an empty line. It is very important that you keep using this format.
Please also make sure that there's always a minimum 1 second gap between the AD segments. The end timestamps should not exceed the video duration.

For example, the following is NOT allowed since there's no gap between the two segments:
=====
0 min 53 sec to 0 min 57 sec
A framed photo of a smiling girl. The title "PAREIDOLIA" appears.

0 min 57 sec to 1 min 0 sec
Credits roll over the photo.
=====

Audio description gaps are determined when there is no dialogue. The recommended approach for audio description is to fit descriptions into the natural pauses/gaps of the original audio. Each gap should be **at least 3 seconds**. The number of words should be depend on the length of the AD gap, one second should allow 3 words maximum, for example, a 3-second gap should contain (3 * 3 =) 9 words approximately. What AD might cover are the following aspects of the visual:
1. form: characters, places, text or any other shape or object.
2. motion: action, time or anything that moves or is indicated by movement.
3. color: including the skin color of characters.
4. sound: visual sound, i.e. sound that is identified only visually.
5. (camera) perspective: bird's-eye view, zoom, point of view, special effects.
6. supportive information: additional information, shifted information.


# Output format
Return as a response a JSON object with the following fields.
- Command: the original command
- TextResponse: response that is read back to the user, describing any actions that the system will take.
- DidChangeTimestamp: true or false. True if the system should jump to a new video timestamp. If true, the NewTimeStamp field must be set.
- NewTimeStamp: timestamp (in number of seconds, e.g., 94) to move the video to.
- DidChangeScript: true or false. True if the AD script is changed. If so, NewScript must be specified.
- NewScript: if DidChangeScript is true, the new Changed AD script (please refer to the format above; if somehow the data isn't exactly in that format, you should make an effort to fix it); otherwise, an empty string.
- DidChangeADLineNumber: true or false. True if the current line number has changed.
- ADLineNumber: Line to jump to in the AD script. If the user requested going to a certain point in the script, move them there and indicate so in the text response. If AD before the user's cursor has changed, you may move the current line so that the user is still at the same location in the script. Note that the AD script may very well contain empty lines, which SHOULD NOT be skipped when counting new lines, in other words, you should treat empty lines in the AD script as a legit line.

# Commands
Here are some possible actions that the user might take. You should identify which action the user is intending based on the prompt.

- Navigate to a specific point in the video using a timestamp (e.g., "go to 1 minute") or description of a specific scene of the video ("go to the scene with the birthday party"). This request may be relative to the current video position ("jump forward ten seconds" or "jump to the next scene").
- Ask a question about the video overall, the current frame, or the current scene. Note that the user may themselves be blind or visually impaired; do not assume that they can see video.
- Generate AD for the video. If the user requests AD, either for the entire video or for a specific part of the video, you should generate the AD, add it to the script in the appropriate location, and return the updated script. In general, you should position the AD where the user's cursor is in the video unless specified otherwise. When possible, try to create AD that does not overlap with speech in the video. Assume that AD will be read out at 100 words per minute.
- Edit or move the current AD segment. This may include editing the text or changing the timestamps of the video.
- Remove AD.
- Edit the entire script in some way, e.g. changing metric units to imperial or renaming a character in the script. In this case, return the updated script and summarize the changes in the text response.
- Find areas in the video that require AD. In this case, seek areas where there are gaps with no speech or AD. You should jump to that part of the video, move the cursor to the appropriate place in the video, and describe what you did in the text response.

The user may enter any command they wish; however there may be some commands that the system cannot perform. For example, the system cannot edit the videos themselves. In that case, the system should respond with a message saying that the action
cannot be performed.

# Associated information
- Conversation history (least recent to most recent): {{conversationHistory}}
- URL of the video: {{videoURL}}
- Current timestamp of the video: {{timestamp}}
- Current line of AD script: {{adScriptLine}}
- Current AD script:
===== START OF AD SCRIPT (NOT A LINE IN THE SCRIPT) =====
{{adScriptText}}
===== END OF AD SCRIPT (NOT A LINE IN THE SCRIPT) =====

\end{lstlisting}


\end{document}